\documentclass[12pt,aps,nofootinbib]{revtex4}

\usepackage{graphicx}
\usepackage{bbm}
\usepackage{amssymb}
\usepackage{amsmath}

\newcommand{\nc}{\newcommand}
\nc{\be}{\begin{equation}}
\nc{\ee}{\end{equation}}
\nc{\bea}{\begin{eqnarray}}
\nc{\eea}{\end{eqnarray}}
\nc{\nn}{\nonumber}
\nc{\sign}{\textrm{sign}}
\nc{\tr}{\textrm{Tr}}
\nc{\slashed}{\slash\hskip -0.2 cm}

\begin{document}

\rightline{UAB-FT-657}
\rightline{NPAC-08-22}

\vskip 1 cm

\title{Separation of Equilibration Time-Scales
in the Gradient Expansion}

\author{Bj\"orn Garbrecht$^{(1)}$}
\author
{Thomas Konstandin$^{(2)}$}

\email[]{bjorn@physics.wisc.edu}
\email[]{konstand@ifae.es}

\affiliation{$^{(1)}$Department of Physics, University of Wisconsin--Madison,
Madison, WI 53706, USA\\
$^{(2)}$IFAE, Universitat Aut\`onoma de Barcelona, 
E-08193 Bellaterra, Barcelona, Spain}


\begin{abstract}
We study thermalization by applying gradient expansion to the
Kadanoff-Baym equations of the 2PI effective action to two-loop in a
theory with Dirac fermions coupled to scalars.  In addition to those
chemical potentials which equilibrate in the on-shell limit, we
identify modes which are conserved in this approximation, but which
relax when off-shell effects are taken into account. This implies that
chemical equilibration does not require higher loop contributions to
the effective action and is compatible with the gradient expansion. We
explicitly calculate the damping time-scales of both, on- and
off-shell, chemical equilibration rates.  It is shown that off-shell
equilibration is suppressed by the thermal width of the particles in
the plasma, which explains the separation of on- and off-shell
chemical equilibration time-scales.
\end{abstract}

\maketitle

%
%

\section{Introduction}

Kadanoff Baym equations are an effective method for
describing the out-of-equilibrium dynamics of quantum fields.
For many particular models and initial conditions,
thermalization has been demonstrated successfully
from numerical solutions to these
equations, see e.g. refs.~\cite{Berges:2000ur, Berges:2001fi,
Berges:2002wr, Aarts:2001qa, Juchem:2003bi, Juchem:2004cs, Berges:2004ce,
Arrizabalaga:2005tf, Berges:2005ai, Arrizabalaga:2005jk, Berges:2005md, 
Lindner:2007am, Tranberg:2008ae} for
some works in the field. While details can depend on the
particular settings, it has been
revealed that starting from far-from-equilibrium
initial conditions, thermalization proceeds typically in several
usually overlapping stages:

\begin{itemize}
\item Dephasing/prethermalization~\cite{Cooper:1996ii, Berges:2004ce}: 
In this stage, a constant equation of state is reached and an
approximate equipartition between kinetic and potential energy is
attained. This dephasing effect does not rely on
scattering-driven processes and is a pure quantum phenomenon.

\item Kinetic equilibration~\cite{Juchem:2003bi, Berges:2005ai, 
Lindner:2007am}: 
Within this stage, the quasi-particle distribution functions approach
the Bose-Einstein or Fermi-Dirac form, respectively. The interactions
which drive the system towards kinetic equilibrium do not necessarily
relax chemical potentials through violation of particle number. For example,
this is the case for elastic two-by-two scatterings.
 
\item Chemical equilibration~\cite{Juchem:2003bi, Lindner:2007am}:
Finally, processes that do not conserve particle numbers
eventually relax the chemical potentials. 
\end{itemize}

This last stage of chemical equilibration is the subject of the present study.
The inclusion of
off-shell effects constitutes one qualitative difference between the Kadanoff-Baym
and the Boltzmann equations. This has been mentioned as a possible explanation
for the discrepancy between numerical solutions within the two approaches
~\cite{Aarts:2001qa}. Besides, Kadanoff-Baym equations include memory
integrals and treat deviations from equilibrium to all order in derivatives,
usually referred to as gradients.
Boltzmann equations do not incorporate memory integrals and are first order
in gradients.

For systems which are close to equilibrium, gradient expansion and
neglect of memory integrals is justifiable. On the other hand, due
to the possible presence of quantities which only equilibrate when
off-shell effects are included, it is desirable to account for these
within an analytical approach.  For the case of scalar $\lambda
\phi^4$ theory in $2+1$ dimensions, Boltzmann equations have been
generalized to include off-shell processes.  Numerical solutions to
these equations are found to be in accordance with the solutions to
the full Kadanoff-Baym equations, but in disagreement with the
on-shell Boltzmann
equations~\cite{Juchem:2003bi,Juchem:2004cs,Cassing:2008nn}. The
explanation is that while $2 \leftrightarrow 2$ scatterings lead to
kinetic equilibration at a time-scale proportional to $\lambda^2$,
chemical equilibration is driven by $1\leftrightarrow 3$ off-shell
processes that are suppressed by higher powers in $\lambda$; see also
e.g. ref.~\cite{Berges:2005ai} for related discussions and
refs.~\cite{Arrizabalaga:2005tf, Arrizabalaga:2005jk} for a numerical
analysis in 3+1 dimensions. The case of an interacting system of
fermions and scalars is numerically treated in~\cite{Lindner:2007am},
where solutions to the Kadanoff-Baym and the Boltzmann equations are
compared. It is found that there are chemical potentials which are
conserved by the Boltzmann equations but which are relaxed when
Kadanoff-Baym equations are applied. This phenomenon is explained by
the fact that the quasi-particle approximation of Boltzmann equations
leads to spuriously conserved quantities.

The  aim of  the current  work is  to show  how to  reproduce  on- and
off-shell chemical equilibration in a semi-classical approach, without
resorting to numerical solutions.  As  a concrete model, we consider a
Dirac fermion  coupled via a Yukawa  coupling to a  complex scalar. We
expect  that  this  model   shares  the  essential  features  for  the
discussion  of   chemical  equilibration  in   the  quark-meson  model
considered  in   refs.~\cite{Berges:2002wr,  Lindner:2007am}  and  the
models usually  considered for leptogenesis~\cite{Fukugita:1986hr} and
for  chargino  mediated  electroweak  baryogenesis~\cite{Huet:1995sh}.
The   classical  Boltzmann   equations   can  be   derived  from   the
Kadanoff-Baym equations by using several approximations. These are the
gradient  expansion  (to  first  order),  the  expansion  in  coupling
constants (typically  up to the  first nontrivial order  that accounts
for  scattering  or  decay   processes)  and  the  quasi-particle  (or
on-shell)  approximation.  We will  show  that  by  omitting the  last
approximation,  full chemical   equilibration  can  be  described  by
semi-classical  transport  equations.   In  particular,  the  gradient
expansion is valid in this  regime and we present analytic expressions
for  the on-  and  off-shell relaxation  time-scales  of the  chemical
potentials.

The paper is organized as follows. In the next section we set up
notation and in section~\ref{sec_GE} we introduce the gradient
expansion. In the two subsequent sections, the model and the Boltzmann
equations in the on-shell limit are discussed. Finally, we calculate
the various time-scales of chemical equilibration in
section~\ref{sec_time_scales}, before we conclude in
section~\ref{sec_concl}.

\section{The Kadanoff-Baym equations\label{sec_KB}}

In this section, we derive the Kadanoff-Baym equations and set up
notation. We follow the nomenclature of ref.~\cite{dickwerk}, where a
more complete treatment can be found.  Statistical systems in
equilibrium and non-equilibrium can be described by the Kadanoff-Baym
equations which are the Schwinger-Dyson equations in the in-in
formalism and in Wigner space~\cite{Calzetta:1986cq}. The
in-in-formalism is designed to describe the time-evolution of density
matrix elements rather than of scattering processes to which the 
usual in-out formalism applies. The in-in formalism is implemented by
integrating along the Schwinger-Keldysh~\cite{Schwinger:1960qe,
Keldysh:1964ud} closed time path (CTP) from a finite time $t=t_0$ to
$t=+\infty$ and back. The Green functions for bosons and fermions are
then defined as the path ordered operators
\bea
{\rm i}\Delta(u,v) &=& 
\left< \Omega \left|  
T_{\cal C} \phi(u) \phi^\dagger(v) \right| 
\Omega \right>, \\
{\rm i}S_{\alpha\beta}(u,v) &=& 
\left< \Omega \left|  
T_{\cal C} \psi_\alpha(u) \bar \psi_\beta(v) 
\right| \Omega \right>. 
\eea
Depending on the locations of the coordinates $u$ and $v$ on the branches
of the CTP, we obtain the four Green functions
\begin{subequations}
\label{Green:Functions}
\bea
{\rm i}\Delta^t(u,v) &=& \left< \Omega \left|  
T  \phi(u) \phi^\dagger(v) \right| 
\Omega \right>, \\
{\rm i}\Delta^<(u,v) &=& \left< \Omega \left|  
\phi^\dagger(v)   \phi(u) \right| 
\Omega \right>, \\
{\rm i}\Delta^>(u,v) &=& \left< \Omega \left|  
  \phi(u) \phi^\dagger(v) \right| 
\Omega \right>, \\
{\rm i}\Delta^{\bar t}(u,v) &=& \left< \Omega \left|  
\bar T  \phi(u) \phi^\dagger(v) \right| 
\Omega \right>, 
\eea
\end{subequations}
and similar expressions for the fermions. In the
following, we focus on the bosonic functions. The fermionic
functions fulfill similar corresponding equations (see
ref.~\cite{dickwerk}). From the definitions~(\ref{Green:Functions}),
it is obvious that only
two of the four functions are independent while the other two can be
expressed as
\begin{subequations}
\bea
\Delta^t(u,v)        &=& \theta(u_0-v_0) \Delta^>(u,v) + 
\theta(v_0-u_0) \Delta^<(u,v), \\
\Delta^{\bar t}(u,v) &=& \theta(u_0-v_0) \Delta^<(u,v) + 
\theta(v_0-u_0) \Delta^>(u,v). 
\eea
\end{subequations}
Besides, the Green functions have the following hermiticity property
\be
\left( {\rm i}\Delta^{<,>}(u,v)\right)^\dagger = {\rm i}\Delta^{<,>}(v,u).
\ee

The retarded and advanced propagators are defined as 
\begin{subequations}
\bea
\Delta^r &=& \Delta^t - \Delta^< = \Delta^> - \Delta^{\bar t}, \\
\Delta^a &=& \Delta^t - \Delta^> = \Delta^< - \Delta^{\bar t}, 
\eea
\end{subequations}
which can be used to define the spectral and hermitian functions
\bea
{\cal A}_\phi &=& \frac{\rm i}2 ( \Delta^r - \Delta^a )
= \frac{\rm i}2 ( \Delta^> - \Delta^< ), \\
\Delta_h &=& \frac12 ( \Delta^r + \Delta^a ) =-{\rm i}\,\sign(u_0 - v_0) 
{\cal A}_\phi. 
\eea
The usefulness of these definitions becomes obvious for the thermal
equilibrium Green functions in Wigner space, where one obtains for a
free boson of mass $M$
\be
\label{spectral_tree}
{\cal A}_\phi(p,x) = \int d^4p \, {\rm e}^{ {\rm i}p \cdot r} 
{\cal A}_\phi(x + r/2, x- r/2) = 
 \pi \delta(k^2 - M^2) \, \sign(k_0),
\ee
and
\be
{\rm i}\Delta^< = 2n_{\rm B}^{\rm eq}(k_0) \, {\cal A}_\phi,\quad
{\rm i}\Delta^> = 2(n_{\rm B}^{\rm eq}(k_0)+1) \, {\cal A}_\phi,\quad
n_{\rm B}^{\rm eq}(k_0) = \frac1{{\rm e}^{\beta k_0} - 1}.
\ee
Using this notation, the Kadanoff-Baym equations for the bosons in
Wigner space read~\cite{dickwerk}
\bea
(\Omega_\phi^2 \pm {\rm i}\Gamma_\phi) * \Delta^{r,a} &=& 1, \label{KB1} \\
\Omega_\phi^2 * \Delta^{<,>} - 
\Pi^{<,>} * \Delta_h &=& \frac12( \Pi^> * \Delta^< -  \Pi^< * \Delta^>),
\label{KB2}
\eea
where
\be
\label{Moyal}
A * B = A \,  {\rm e}^{-\frac{\rm i}2 
(\overleftarrow \partial_x \overrightarrow \partial_k  
 - \overleftarrow \partial_k \overrightarrow \partial_x  ) } \, B
\ee
denotes the Moyal star product and the various functions $\Pi$ are
self-energies which have the same time-ordering prescriptions as
defined for the Green functions in eqs.~(\ref{Green:Functions}). The
self-energies can be deduced from the two-particle-irreducible (2PI)
effective action.  Besides, we have defined
\be
\Omega_\phi^2 = k^2 - M^2 - \Pi_h, \quad \Gamma_\phi=\frac{\rm i}2(\Pi^>-\Pi^<).
\ee
In the following, we discuss this system of equations in the
gradient expansion.

\section{Gradient Expansion\label{sec_GE}}

We consider spatially homogeneous and isotropic systems, such that all
functions depend on the time $t=x_0$, but not on the spatial
coordinates $\vec x$. At rather late times, when the system is not too
far from equilibrium, one expects that the relaxation of the system
towards equilibrium is driven by interactions which allow for a
gradient expansion in the weak coupling regime.\footnote{This argument
can however be jeopardized by oscillation effects in the case of
several flavors as discussed in ref.~\cite{Konstandin:2004gy,
Konstandin:2005cd}.} Thus, one can expand the Moyal star
product~(\ref{Moyal})
\be
A * B \approx A\, B - \frac{\rm i}2\{ A, B\}_{pb} + O(\partial_x^2),
\ee
where we have defined the Poisson brackets $\{A,B\}_{pb} = \partial_x A
\partial_k B- \partial_k A \partial_x B$. 

At first order in the gradient expansion, eq.~(\ref{KB1}) reads
\be
(\Omega_\phi^2 \pm {\rm i} \Gamma_\phi) \Delta^{r,a} - 
\frac{\rm i}2 \{ \Omega_\phi^2 \pm {\rm i}\Gamma_\phi , \Delta^{r,a} \}_{pb}  = 1,
\ee
which is solved by 
\be
\Delta^{r,a} = \frac{1}{\Omega_\phi^2 \pm {\rm i}\Gamma}.
\ee
Here, we have used the fact that $\{ A , 1/A\}_{pb} = -\{ A , A\}_{pb}/A^2 =0 $. This
leads to the spectral and hermitian functions
\be
{\cal A}_\phi = \frac{\Gamma_\phi}{\Omega_\phi^4 + \Gamma_\phi^2}, \quad
\Delta_h = \frac{\Omega_\phi^2}{\Omega_\phi^4 + \Gamma_\phi^2}.
\ee
The remaining information is encoded in the anti-hermitian part of
eq.~(\ref{KB2}), which reads
\be
- \{ \Omega^2 , {\rm i}\Delta^{<,>}\}_{pb} 
+ \{ {\rm i}\Pi^{<,>} , {\rm i}\Delta_h \}_{pb}
= \Pi^> \Delta^< - \Pi^< \Delta^>.
\ee
Since we assume that the coupling constant $y$ is small, the
self-energies are at late times effectively of first order in the
gradient expansion. We therefore neglect the self-energies in the
Poisson brackets (in contrast to the methods used e.g.~in
ref.~\cite{Cassing:2008nn}), which leads to
\be
\label{eq_kin_bos}
2 k_0 \partial_t {\rm i}\Delta^{<,>}
= \Pi^> \Delta^< - \Pi^< \Delta^>.
\ee
Analogously, one finds for a free fermion in equilibrium 
\be
{\rm i} S^< = -2n_{\rm F}^{\rm eq}(k_0) \, {\cal A}_\psi,\quad
{\rm i} S^> = 2(1-n_{\rm F}^{\rm eq}(k_0)) \, {\cal A}_\psi,\quad
n_{\rm F}^{\rm eq}(k_0) = \frac1{{\rm e}^{\beta k_0} + 1},
\ee
and the fermionic Kadanoff-Baym equations 
\bea
\left( \slashed \Omega_\psi \pm {\rm i}\slashed \Sigma_{\cal A} \right) 
S^{r,a} 
- \frac{\rm i}2 \left\{ \slashed \Omega_\psi \pm {\rm i}\slashed \Sigma_{\cal A} , 
S^{r,a} \right\}_{pb}
&=& 1,
\\
\label{eq_kin_fer}
\gamma_0 \partial_t {\rm i}S^{<,>} &=& \frac12(\slashed \Sigma^> S^< - \slashed \Sigma^< S^>) + \, {\rm h.c.},
\eea
where $\slashed\Sigma$ denotes the fermionic self-energy and $\slashed \Omega_\psi =\slashed p - \slashed \Sigma_h$ and $\slashed
\Sigma_{\cal A} = \frac{\rm i}2 (\slashed \Sigma^> - \slashed
\Sigma^<)$. This implies that
\bea
\label{eq_spec_ferm}
{\cal A}_\psi &=& 
\frac{\slashed \Omega_\psi \Gamma_\psi 
- \slashed \Sigma_{\cal A} (\Omega_\psi^2 - \Sigma_{\cal A}^2)}
{( \Omega_\psi^2 - \Sigma_{\cal A}^2)^2+ \Gamma_\psi^2}, \\
S_h &=& 
\frac{\slashed \Omega_\psi (\Omega_\psi^2 - \Sigma_{\cal A}^2)
+ \slashed \Sigma_{\cal A}  \Gamma_\psi}
{( \Omega_\psi^2 - \Sigma_{\cal A}^2)^2 + \Gamma_\psi^2},
\eea
where we have defined $\Gamma_\psi = 2 \Omega_\psi \cdot
\Sigma_\psi=\{\slashed\Sigma_{\cal A},\slashed\Omega_\psi\}$,
$\Sigma_{\cal A}^2=\slashed\Sigma_{\cal A}\slashed\Sigma_{\cal A}$ and
$\Omega_\psi^2= \slashed\Omega_\psi\slashed\Omega_\psi$. In general,
the fermionic propagator has a complicated structure and e.g. contains
a pole from a collective excitation~\cite{Weldon:1989ys}. However,
the corresponding residuum is exponentially small for most particles
in the plasma such that one can safely ignore this subtlety.

\section{The Model and its Conserved Charges\label{sec_model}} 

For our purpose of an analytic calculation of the separation
of time-scales of on- and off-shell chemical equilibration, 
we consider a theory with a Lagrange density as in ref.~\cite{dickwerk}
\be
\label{Lagrangian}
{\cal L} = {\rm i}\bar \psi \slashed\partial \psi 
+ \partial_\mu \phi^\dagger \partial^\mu \phi - \phi^\dagger M^2 \phi
- y \, \bar \psi (P_R \phi + P_L \phi^\dagger ) \psi,
\ee
with a small real coupling constant $y$. Using the two-loop 2PI
effective action, this Lagrangian leads to the following expressions
for the self-energies~\cite{dickwerk}
\be
\label{selferg:bose}
{\rm i}\Pi^{<,>} (k,x) = - y^2 \int \frac{d^4p d^4q}{(2\pi)^4} 
\delta(k-p-q) \tr[P_R {\rm i}S^{>,<} (-p,x) P_L {\rm i}S^{<,>} (q,x) ], 
\ee
and for the fermions ($\slashed\Sigma = \slashed\Sigma_L + \slashed\Sigma_R$) 
\begin{subequations}
\label{selferg:fermi}
\bea
{\rm i}\slashed \Sigma_R^{<,>} (p,x) &=& y^2 \int \frac{d^4q d^4k}{(2\pi)^4} 
\delta(k-p-q) {\rm i}\Delta^{>,<} (-k,x) P_L {\rm i}S^{<,>} (-q,x) P_R, \\
{\rm i}\slashed \Sigma_L^{<,>} (q,x) &=& y^2 \int \frac{d^4p d^4k}{(2\pi)^4} 
\delta(k-p-q) {\rm i}\Delta^{<,>} (k,x) P_R {\rm i}S^{<,>} (-p,x) P_L.
\eea
\end{subequations}

We now define the charges
\begin{subequations}
\label{def_conservs}
\bea
Q_S &=&  \int \frac{d^4k}{(2\pi)^4} 2 k_0 
({\rm i}\Delta^< + 2\theta(-k_0) {\cal A}_\phi) ,  \\ 
P^\mu_S &=&  \int \frac{d^4k}{(2\pi)^4} 2 k^\mu k_0 
({\rm i}\Delta^< + 2\theta(-k_0) {\cal A}_\phi),  \\
Q_{L/R} &=&  -\int \frac{d^4q}{(2\pi)^4} \tr \gamma_0  P_{L/R} 
( {\rm i}S^< + 2\theta(-q_0) {\cal A}_\psi) ,  \\
P^\mu_{L/R} &=&  -\int \frac{d^4q}{(2\pi)^4} \tr  q^\mu \gamma_0 P_{L/R} 
( {\rm i}S^< + 2\theta(-q_0) {\cal A}_\psi),
\eea
\end{subequations}
where the terms involving the spectral functions ${\cal A}$ have
vanishing time-derivatives and have been introduced to make the
expressions finite. The interpretation of these quantities in
the on-shell limit is given in the next section.
When using the
kinetic equations~(\ref{eq_kin_bos}) and~(\ref{eq_kin_fer}),
the form of the collision terms immediately leads to the following
conserved quantities:
\begin{subequations}
\label{eq_conservs}
\bea
\label{Q:chiral}
Q_R + Q_L &=& {\rm const.}  \\
\label{Q:Yukawa}
2Q_S - Q_R + Q_L &=& {\rm const.}  \\
P^\mu_S + P^\mu_R + P^\mu_L &=& {\rm const.}
\eea
\end{subequations}
We emphasize that these conservation laws are due to the symmetries of
the collision terms and make no assumption about the Green functions and their
spectral properties. Of course, this is reflecting that above charges
are conserved due to Noether symmetries with respect to rephasings and
translations of the Lagrangian~(\ref{Lagrangian}).

\section{The On-Shell Limit\label{sec_on_shell}}

The on-shell limit arises when neglecting the influence of
interactions on the spectral functions. Notice, that in order to
obtain the correct results, one has to take into account that the
Breit-Wigner width behaves as $\Gamma \to \sign(k_0) \epsilon$ for
both bosons and fermions. In this case, we recover the spectral functions of the free theory
\bea
{\cal A}_\phi(k,x) &\to& 
 \pi \delta(k^2 - M^2) \, \sign(k_0), \\
 {\cal A}_\psi(k,x) &\to& 
 \pi \slashed k \delta(k^2) \, \sign(k_0),
\eea
that fulfill
\bea
\int \frac{dk_0}{2\pi} \, 2k_0 {\cal A}_\phi &=& 1, \nn \\
\int \frac{dk_0}{2\pi} \, 2\gamma_0 {\cal A}_\psi &=& \mathbbm{1}.
\eea
Next, we employ the ansatz
\begin{subequations}
\label{propagators}
\bea
{\rm i}\Delta^< &=& 2n_S(k_0) \, {\cal A}_\phi\,,\\
{\rm i}\Delta^> &=& 2(n_S(k_0)+1) \, {\cal A}_\phi\,, \\
{\rm i}S^< &=& -2(n_L(k_0) P_L  {\cal A}_\psi P_R + n_R(k_0) P_R {\cal A}_\psi P_L ) \,, \\
{\rm i}S^> &=& 2({\cal A}_\psi - n_L(k_0) P_L {\cal A}_\psi P_R - n_R(k_0) P_R {\cal A}_\psi P_L) \, ,
\eea
\end{subequations}
with the generalized particle distribution functions $n(k_0)$. This
turns the conserved quantities in eq. (\ref{def_conservs}) into
\bea
Q_S &=& \int \frac{d^3k}{(2\pi)^3} 
(n_S(\omega) - \bar n_S(\omega)), \\
P^0_S &=& \int \frac{d^3k}{(2\pi)^3} 
(\omega n_S(\omega) + \omega \bar n_S(\omega)), \\
Q_{L,R} &=& \int \frac{d^3k}{(2\pi)^3} 
(n_{L,R}( k ) - \bar n_{L,R}( k )), \\
P^0_{L,R} &=& \int \frac{d^3k}{(2\pi)^3} 
(k n_{L,R}(k) + k \bar n_{L,R}(k)), 
\eea
where we have introduced the bosonic dispersion relation $\omega^2 =
\mathbf{k}^2 + M^2$. Besides, we have defined the anti-particle
distribution functions for negative energies
\be
\bar n_S(\omega) = - (n_S(-\omega)+1), \quad
\bar n_{L,R}(k_0) =  1 - n_{L,R}(-k_0).
\ee

\begin{figure}
\begin{center}
\includegraphics*[width=0.6 \textwidth]{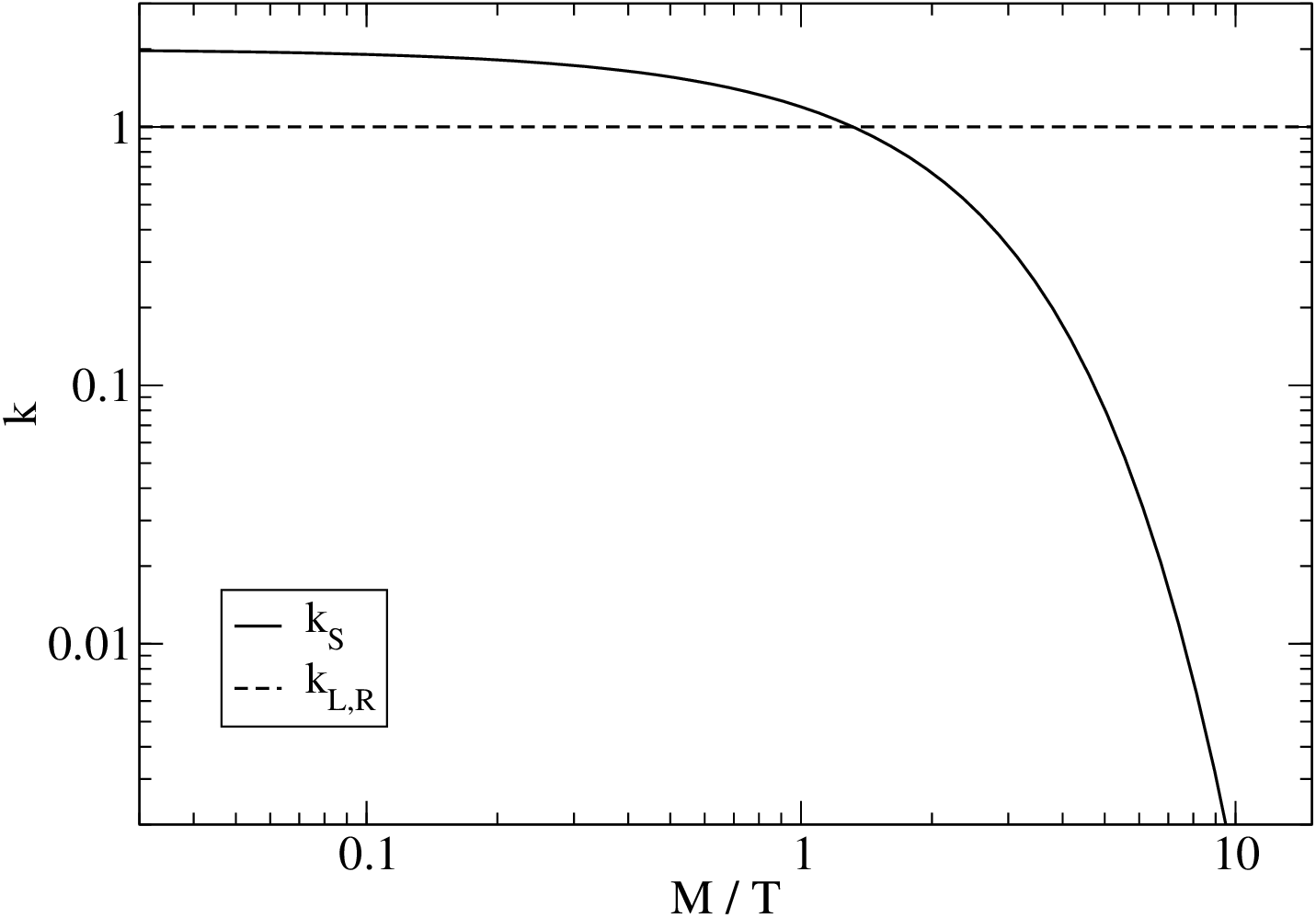}
\end{center}
\vskip -0.5cm
\caption{
The plots show the coefficients $k_{L,R}$ and $k_S$ as functions of
the scalar mass $M$.}
\label{fig_kappas}
\end{figure}

From these definitions, it is clear that the conserved quantities
represent total energy and charges. In particular, the knowledge of the
initial charges and total energy should suffice to determine the final
equilibrium state of the system that is characterized by the temperature and the
chemical potentials after equilibration. Assuming kinetic equilibrium, the
Kubo-Martin-Schwinger (KMS) relation including chemical potentials
are
\be
\label{KMS}
\Delta^> = {\rm e}^{\beta(p_0-\mu_S)} \Delta^<, \quad
P_L S^> =  -{\rm e}^{\beta(p_0-\mu_L)}  P_L S^<, \quad
P_R S^> =  -{\rm e}^{\beta(p_0-\mu_R)}  P_R S^<. \quad
\ee
These relations then imply that the distribution functions are of the form
\be
n_{L,R}=1/({\rm e}^{\beta(p_0-\mu_{L,R})}+1)\,,\quad
n_S=1/({\rm e}^{\beta(p_0-\mu_S)}-1)\,.
\ee
Under the close-to-equilibrium assumption that $\mu_{L,R,S}/T\ll 1$, we can
linearly relate
the charges to the chemical potentials by introducing statistical factors as
\be
Q_{L,R,S}=\frac{T^2}{6}k_{L,R,S}\mu_{L,R,S}\,,
\ee
where
\bea
k_{L,R}\!\!\!&=&\!\!\!12T^{-3}\int\frac{d^3k}{(2\pi^3)}n_{\rm F}^{\rm eq}(1-n_{\rm F}^{\rm eq})
=\frac6{\pi^2}\int\limits_{0}^{\infty}x^2dx\frac{{\rm{e}^x}}{({\rm e}^x+1)^2}
=1
\,,\\
k_{S}\!\!\!&=&\!\!\!12T^{-3}\int\frac{d^3k}{(2\pi^3)}n_{\rm B}^{\rm eq}(n_{\rm B}^{\rm eq}+1)
=\frac6{\pi^2}\int\limits_{M/T}^{\infty}x\,dx
\sqrt{x^2-\frac{M^2}{T^2}}
\frac{{\rm{e}^x}}{({\rm e}^x-1)^2}
\,.
\eea
In the limit $M/T\to0$, the latter integral evaluates to $k_S=2$, and
we plot the statistical factors as a function of $M/T$ in
Fig~\ref{fig_kappas}.

Next, we note that the self-energies~(\ref{selferg:bose})
and~(\ref{selferg:fermi}) inherit the KMS property~(\ref{KMS}) from
the Wightman functions,
\begin{subequations}
\bea
\Pi^>(k)&=&\Pi^<(k){\rm e}^{\beta(k_0-\mu_L+\mu_R)},
\\
\slashed \Sigma^>_R(k)&=&-\slashed \Sigma_R^<(k){\rm e}^{\beta(k_0-\mu_L+\mu_S)},
\\
\slashed \Sigma^>_L(k)&=&-\slashed \Sigma_L^<(k){\rm e}^{\beta(k_0-\mu_R-\mu_S)}.
\eea
\end{subequations}
From the Feynman rules in the Schwinger-Keldysh formalism (in
particular from energy-momentum conservation) it is possible to
derive that these relations even hold to arbitrary order in
perturbation theory. Consequently, we find for the collision terms
\begin{subequations}
\bea
\Pi^>(k)\Delta^<(k)-\Pi^<(k)\Delta^>(k)&=&
\Pi^>(k)\Delta^<(k)\left(1-{\rm e}^{\beta(\mu_L-\mu_R-\mu_S)}\right),
\\
\slashed\Sigma_L^>(k) S_L^<(k)-\slashed\Sigma_L^<(k)S_L^>(k)&=&
\slashed\Sigma_L^>(k) S_L^<(k)\left(1-{\rm e}^{\beta(\mu_R-\mu_L+\mu_S)}\right),
\\
\slashed\Sigma_R^>(k) S_R^<(k)-\slashed\Sigma_R^<(k)S_R^>(k)&=&
\slashed\Sigma_R^>(k) S_R^<(k)\left(1-{\rm e}^{\beta(\mu_L-\mu_R-\mu_S)}\right),
\eea
\end{subequations}
such that these terms vanish and a static solution of the transport
equations in eqs.~(\ref{eq_kin_bos}) and~(\ref{eq_kin_fer}) is
obtained if
\be
\label{chemeq}
\mu_S + \mu_R - \mu_L = 0.
\ee
Hence, one would expect that the final state is determined by the temperature
and the three conserved quantities in
eq.~(\ref{eq_conservs}) only, and that it satisfies the equilibrium
condition~(\ref{chemeq}) above.

However, it turns out that in the on-shell approximation, not only
the combinations~(\ref{eq_conservs}) are conserved, but also
additional charges in the particle and
anti-particle sectors separately~\cite{Lindner:2007am}. This is a
purely kinematic effect. Consider a scalar decaying into a
fermion/anti-fermion pair. If all three particles are on-shell, all
energies have to have the same sign. This is most easily seen in the
rest-frame of the scalar, where $k_0 = \pm M$. In this frame, the
momenta of fermions differ only by a sign and hence their energies are
equal and therefore $q_0=p_0=k_0/2$. This means that the collision
integrals in eqs.~(\ref{eq_kin_bos}) and (\ref{eq_kin_fer}) have in
the on-shell limit only a support in the regions
\be
\label{support}
\sign{k_0} = \sign{p_0} = \sign{q_0}. 
\ee
This is sufficient to show that the number of fermions
plus twice the number of scalars is separately conserved
for particles and for anti-particles, such that the Boltzmann
equation cannot reproduce chemical equilibration in the on-shell limit.
Formally, we can express these conservation laws by defining
\be
\label{oddcharges}
\bar Q_{L,R,S}=\int\frac{d^3k}{2\pi^3}(n_{L,R,S}(\omega)+\bar n_{L,R,S}(\omega))
\ee
and noting
\begin{subequations}
\label{oddcharges:conserved}
\bea
2\bar Q_S+\bar Q_L +\bar Q_R&=&{\rm const.}\\
\bar Q_L- \bar Q_R&=&{\rm const.}
\eea
\end{subequations}
In order to distinguish from the conserved charges defined in
eqs.~(\ref{eq_conservs}), we refer to these combinations as total
particle number densities. We introduce an additional set of chemical
potentials $\bar\mu_{L,R,S}$ to account for total particle numbers
different from their equilibrium values. These are defined through the
relations
\be
\Delta^> = {\rm e}^{\beta(p_0-\bar \mu_S \sign p_0 )} \Delta^<, \quad
\ee
\be
P_L S^> =  -{\rm e}^{\beta(p_0-\bar \mu_L \sign p_0 )} P_L S^<, \quad
P_R S^> =  -{\rm e}^{\beta(p_0-\bar \mu_R \sign p_0 )} P_R S^<. \quad
\ee
Now by taking combinations of $\mu_{L,R,S}$ and $\bar\mu_{L,R,S}$, we
can adjust the number of particles and anti-particles independently.

For a support of the form as in eq. (\ref{support}), the collision
terms for non-vanishing $\bar\mu_{L,R,S}$ read
\begin{subequations}
\bea
\Pi^>(k)\Delta^<(k)-\Pi^<(k)\Delta^>(k)&=&
\Pi^>(k)\Delta^<(k)\left(1-{\rm e}^{\sign(k_0)\beta(\bar\mu_L+\bar\mu_R-\bar\mu_S)}\right),
\\
\slashed\Sigma_L^>(k) S_L^<(k)-\slashed\Sigma_L^<(k)S_L^>(k)&=&
\slashed\Sigma_L^>(k) S_L^<(k)\left(1-{\rm e}^{\sign(k_0)\beta(-\bar\mu_R-\bar\mu_L+\bar\mu_S)}\right),
\\
\slashed\Sigma_R^>(k) S_R^<(k)-\slashed\Sigma_R^<(k)S_R^>(k)&=&
\slashed\Sigma_R^>(k) S_R^<(k)\left(1-{\rm e}^{\sign(k_0)\beta(-\bar\mu_L-\bar\mu_R+\bar\mu_S)}\right),
\eea
\end{subequations}
and these vanish as long as 
\be
\label{eq_zero_mode}
\bar \mu_S - \bar \mu_L - \bar \mu_R = 0. 
\ee
Notice that this is only true in the two-loop approximation of the
effective action.

To summarize, we have defined six chemical charges, $Q_{L,R,S}$ and
$\bar Q_{L,R,S}$. Through on-shell processes, two linear combinations
of these charges are not conserved and equilibrate according to the
conditions~(\ref{chemeq}) and~(\ref{eq_zero_mode}). Therefore, in the
on-shell limit there remain four conserved chemical charges as defined
by eqs.~(\ref{Q:chiral},~\ref{Q:Yukawa})
and~(\ref{oddcharges:conserved}).  While the
charges~(\ref{Q:chiral},~\ref{Q:Yukawa}) are conserved by virtue of
Lagrangian symmetries, the conservation of the
charges~(\ref{oddcharges:conserved}) is an artifact of the on-shell
approximation.  Notice that this argument is in principle not based on
the gradient expansion, since for these time-independent chemical
potentials the collision terms vanish to all orders in gradients as
long as the support of the collision integral is of the form as in
eq.~(\ref{support}).

\section{Time-Scales\label{sec_time_scales}}

The considerations of the last section show that the relaxation of
some chemical potentials is described by on-shell Boltzmann
equations. The rate of these relaxation processes is given by the
thermally averaged matrix elements of the on-shell scattering
process. Off-shell equilibration on the other hand is additionally
suppressed by the Breit-Wigner width of the particles in the plasma,
which is why we expect these processes to be slow compared to the
on-shell relaxation.  In the following, we analytically calculate the
on-shell relaxation time-scale $\tau_{\rm on}$ and the off-shell
time-scale scale $\tau_{\rm off}$ in linear response theory, and we
show that indeed, they exhibit a separation $\tau_{\rm on}\ll\tau_{\rm
off}$. For this purpose, we assume that the deviations from
equilibrium are of the form
\bea
\delta \Delta^<(k) = \delta \Delta^>(k) &=& 
2 \beta \delta \mu_S(k) n_{\rm B}^{\rm eq}(k) 
(n_{\rm B}^{\rm eq}(k)+1) {\cal A}_\phi, \\
\delta S^<(k) = \delta S^>(k) &=& 
-2 \beta ( P_L \delta \mu_L(k) + P_R \delta \mu_R(k) ) 
n_{\rm F}^{\rm eq}(k) (1 - n_{\rm F}^{\rm eq}(k)) {\cal A}_\psi.
\eea
These expressions follow from the definitions~(\ref{propagators})
with
\be
n_S(k)=\frac1{{\rm e}^{\beta (k_0-\delta\mu_S(k))}-1}\,,\quad
n_{L,R}(k)=\frac1{{\rm e}^{\beta (k_0-\delta\mu_{L,R}(k))}+1}\,,\quad
\ee
when expanded for $\delta\mu_{S,R,L}/T\ll1$.

\subsection{Charge Equilibration}

Let us now calculate the relaxation rate for the charge densities that
is damped by on-shell processes. Therefore, we consider the case of
chemical potentials without energy dependence, $\delta \mu_{L,R,S}=
\mu_{L,R,S}$.

\begin{figure}
\begin{center}
\includegraphics*[width=0.6 \textwidth]{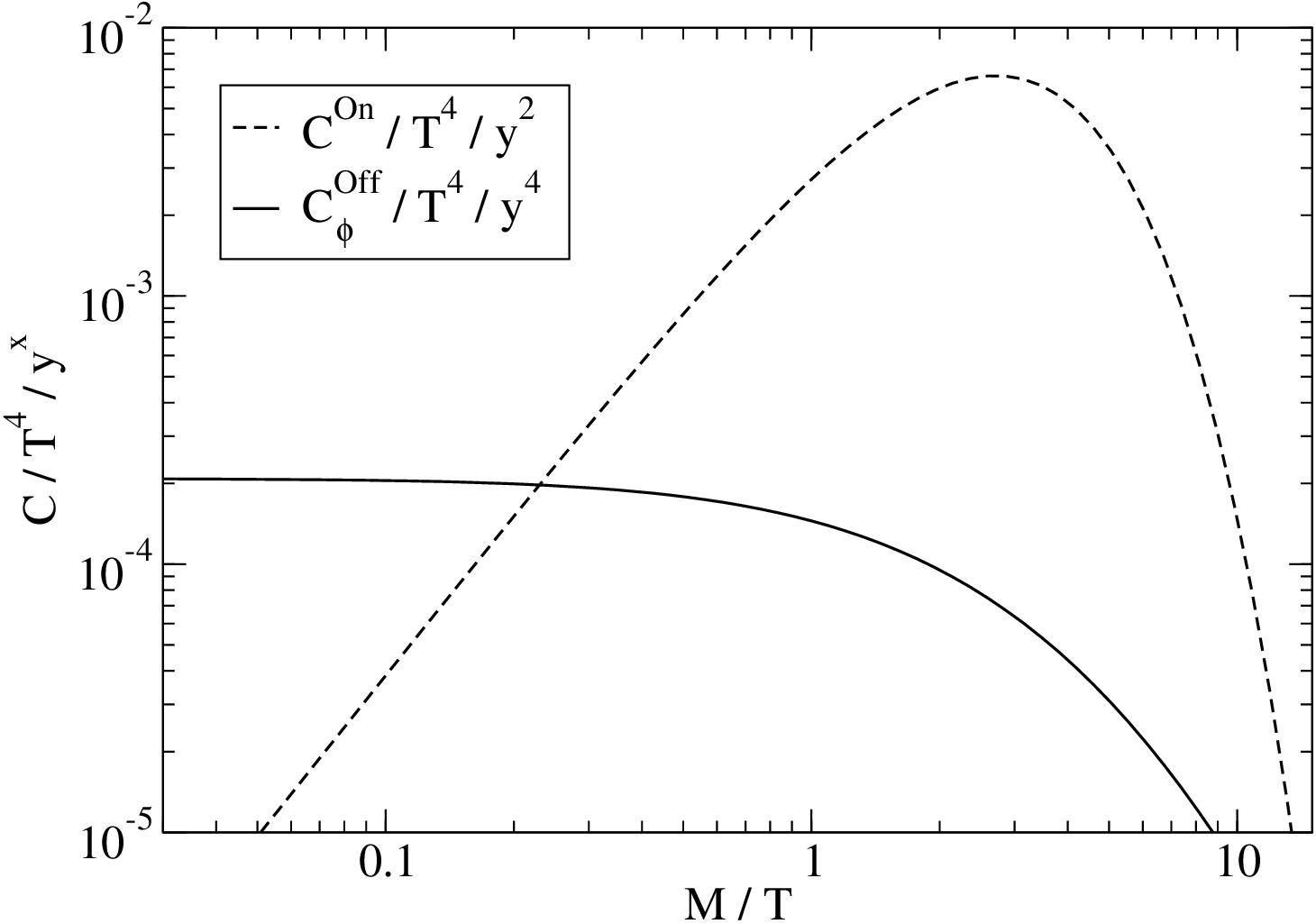}
\end{center}
\vskip -0.5cm
\caption{
The plot shows the collision terms ${\cal C}^{\rm on}$ and ${\cal
C}_\phi^{\rm off}$ as functions of the scalar mass $M$. The
combination ${\cal C}_\phi^{\rm off}/y^4$ has a small residual
dependence on $y$ and we use $y=1.0$ in the plot.}
\label{fig_Cs}
\end{figure}

Using the kinetic equations (\ref{eq_kin_bos}) and (\ref{eq_kin_fer}),
one obtains in linear response three equations
\begin{subequations}
\label{eq_sys_chem}
\bea
\partial_t Q_L &=&  \frac1T {\cal C}^{\rm on} ( \mu_S + \mu_R - \mu_L ), \\
\partial_t Q_R &=&  -\frac1T {\cal C}^{\rm on} ( \mu_S + \mu_R - \mu_L ), \\
\partial_t Q_S &=&  -\frac1T {\cal C}^{\rm on} ( \mu_S + \mu_R - \mu_L ), 
\eea
\end{subequations}
what is consistent with the conserved combinations in
eq.~(\ref{eq_conservs}). The equation for
the non-conserved combination of chemical
potentials~(\ref{chemeq}) is
\be
\partial_t ( \mu_S + \mu_R - \mu_L ) =  
-\frac6{T^3}
\left(\frac2{k_{L,R}} + \frac1{k_S}\right)
\, {\cal C}^{\rm on} ( \mu_S + \mu_R - \mu_L )\,,
\ee
where in on-shell approximation
\bea
{\cal C}^{\rm on} &=&  -y^2 \int \frac{d^4p d^4q d^4k}{(2\pi)^{8}} 
\delta(k-p-q) \tr [P_R iS^> (-p) P_L iS^< (q) ] i\Delta^>(k) ] \\
&=& 2 y^2 \int \frac{d^4p d^4q }{(2\pi)^5} 
 p_\mu q^\mu  n_{\rm F}^{\rm eq}(p_0) n_{\rm F}^{\rm eq}(q_0) (1+n_{\rm B}^{\rm eq}(p_0 + q_0)) \nn \\
&& \times \, \sign(q_0)\delta(q^2) \, \sign(p_0)\delta(p^2) \, 
\sign(q_0+p_0)\delta((q+p)^2-M^2) \,.
\eea
After energy and angular integration, this reads
\bea
\label{C:on}
{\cal C}^{\rm on} &=&  y^2 M^2 \int \frac{dp dq }{16\pi^3} 
  \theta(4q p - M^2)  n_{\rm F}^{\rm eq}(p) n_{\rm F}^{\rm eq}(q) (1 +  n_{\rm B}^{\rm eq}(p + q) )\,.
\eea
The symmetry of this expression becomes more apparent when noting that
\be
n_{\rm F}^{\rm eq}(p) n_{\rm F}^{\rm eq}(q) (1 +  n_{\rm B}^{\rm eq}(p + q) )
=( 1 -  n_{\rm F}^{\rm eq}(p))(1- n_{\rm F}^{\rm eq}(q) )  n_{\rm B}^{\rm eq}(p + q)\,.
\ee
The integral~(\ref{C:on}) has an interpretation in terms of processes that
describe
the decay of the scalar into a fermion--anti-fermion pair and
annihilation of a fermion--anti-fermion pair into a scalar. These
processes are suppressed for small scalar masses, since this reduces
the phase space of the kinetically allowed final states. We plot
${\cal C}^{\rm on}$ in Fig.~\ref{fig_Cs}, from where its features can be verified.

\begin{figure}
\begin{center}
\includegraphics*[width=0.95 \textwidth]{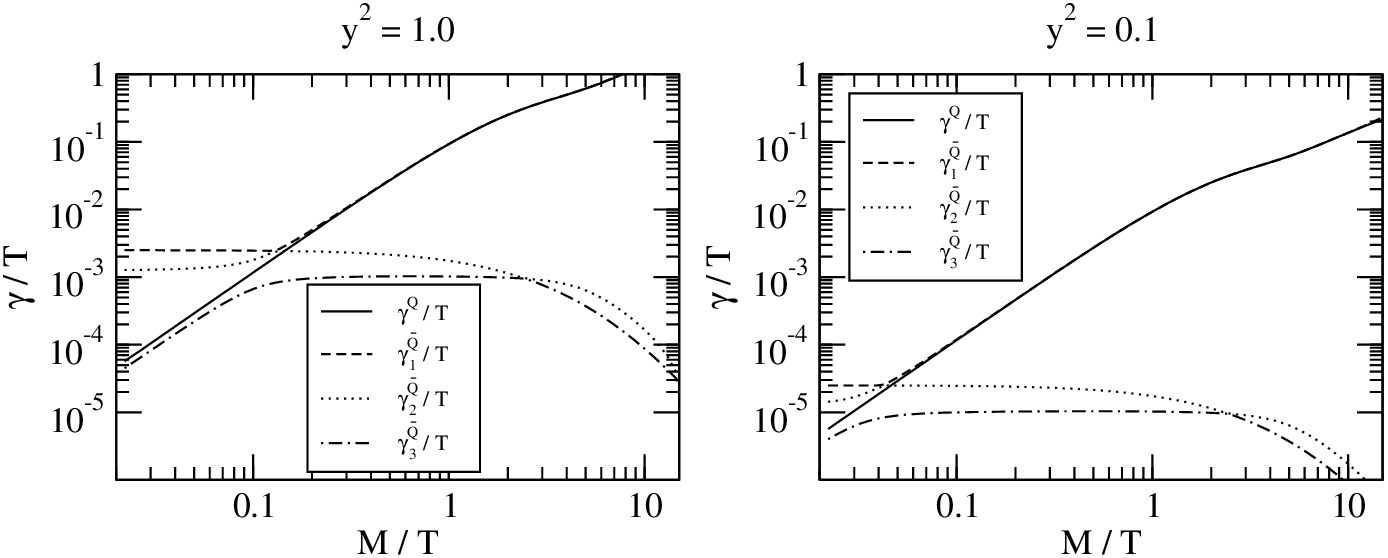}
\end{center}
\vskip -0.5cm
\caption{
The plots show the time-scales of on-shell chemical equilibration,
$\gamma^Q$ given by eq.~(\ref{gamma_on}), and the three different
time-scales given by the eigenvalues of the system in
eq.~(\ref{eq:off-shell}) as a function of the scalar mass $M$. In the
left (right) plot a coupling $y^2=1.0$ ($y^2=0.1$) has been used.}
\label{fig_gammas}
\end{figure}

From there the typical time-scale of charge equilibration can be read
off:
\be
\label{gamma_on}
 \gamma^Q = \frac6{T^3}(k_S^{-1} + 2 k_{L,R}^{-1}) \, {\cal
 C}^{\rm on}.
\ee
The resulting time scale for the choices $y^2=1.0$ and $y^2=0.1$ is
shown in Fig.~\ref{fig_gammas}.

For $M/T\gg1$, it is instructive to evaluate the relevant integrals as
\be
k_S\sim  \sqrt{18/\pi^3} \,
{\rm e}^{-\frac{M}{T}} \left(\frac{M}{T}\right)^{\frac32}
\ee
and
\be
{\cal C^{\rm on}}\sim
\frac{y^2}{\sqrt{ 512 \pi^5}} M^\frac52 T^\frac32 {\rm e}^{-\frac{M}{T}},
\ee
such that
\be
\gamma^Q \sim \frac{M}{16\pi}\,.
\ee
Therefore, chemical equilibration according to~(\ref{chemeq}) is attained
faster for larger $M$. We emphasize however, that this does not imply that
physical interaction rates become fast. The opposite is true, as can be
seen from the exponential decrease of ${\cal C^{\rm on}}$. Chemical equilibration
can occur, because in the large mass limit, a change in
the chemical potential $\mu_S$
only corresponds to an exponentially small change in the physical charge density
$Q_S$. The increase of $\gamma^Q$ with larger $M$ has been noted
in refs.~\cite{Chung:2008aya,Chung:2008gv}. There it has been shown that
this feature simplifies considerations of chemical equilibration
for electroweak baryogenesis and for the conversion of baryon-minus-lepton number
to baryon number, as relevant for leptogenesis.

\subsection{Total Particle Number Equilibration}

In order to investigate the equilibration of total particle number,
consider chemical potentials of the form
\be
\delta \mu_{L,R,S}(k) = \bar \mu_{L,R,S} \, \sign(k_0). 
\ee
Using this ansatz in the kinetic equations (\ref{eq_kin_bos}) and
(\ref{eq_kin_fer}) and linearizing in the chemical potentials then
leads to
\begin{subequations}
\label{eq:off-shell}
\bea
\frac{T^3}6 k_S \partial_t \bar \mu_S &=& - {\cal C}^{\rm on} (\bar \mu_S - 2\bar \mu_+) 
- {\cal C}^{\rm off} \bar \mu_S, \\
\frac{T^3}6 k_{L,R} \partial_t \bar \mu_+ &=& {\cal C}^{\rm on} (\bar \mu_S - 2\bar \mu_+),  \\
\frac{T^3}6\ k_{L,R} \partial_t \bar \mu_- &=& - {\cal C}^{\rm off} \bar \mu_-,
\eea
\end{subequations}
where we have defined the combinations $\bar \mu_\pm = \frac12 ( \bar \mu_L \pm
\bar \mu_R )$ and the off-shell decay rate
\bea
{\cal C}^{\rm off} &=& y^2
\int \frac{d^4p d^4q d^4k}{(2\pi)^8} 
\delta(k-p-q) \tr [P_R iS^> (-p) P_L iS^< (q) ]{\rm i}\Delta^>(k) \nn \\
&& \times  [ \, \sign(k_0) - (\sign(q_0) + \sign(p_0))/2]^2.
\eea

Notice that without the off-shell contributions, there would be two
unsuppressed modes, namely $\bar \mu_-$ and $k_S \bar \mu_S+ k_{L,R}\bar
\mu_+$. On the other hand, including the off-shell effects and in the limit of
${\cal C}^{\rm off}/{\cal C}^{\rm on}\ll 1$, these are still
approximate eigenmodes which are damped at the rates
\be
\gamma^{\bar Q}_1\approx
\frac6{T^3}k_{L,R}^{-1} {\cal C}^{\rm off}\,, \quad
 \gamma^{\bar Q}_2\approx
\frac6{T^3}k_S^{-1} {\cal C}^{\rm off}\,.
\ee
In addition, there is the approximate mode
$ \bar\mu_S- 2 \bar\mu_+$, which is suppressed by
on-shell equilibration at the rate
\be
\gamma^{\bar Q}_3 \approx
\frac6{T^3}(2k_{L,R}^{-1} + k_S^{-1}) {\cal C}^{\rm on}.
\ee

Let us now proceed with the evaluation of the decay rate ${\cal C}^{\rm off}$
close to
equilibrium. The main contributions to this integral come from the
regions, where two of the particles are on-shell, but the third is
off-shell violating the condition in eq. (\ref{support}). We hence
replace two of the Breit-Wigner spectral functions by delta
functions. For simplicity, we present only the contribution where the
scalar is off-shell, in which case one obtains the contribution
\bea
{\cal C}^{\rm off}_\phi &=& y^2  \int \frac{d^4p d^4q }{ 16\pi^6} 
 p_\mu q^\mu  n_{\rm F}^{\rm eq}(p_0) n_{\rm F}^{\rm eq}(q_0) n_{\rm B}^{\rm eq}(-p_0 - q_0) \nn \\
&& \times \, \sign(q_0)\delta(q^2) \, \sign(p_0)\delta(p^2) \, 
\frac{\Gamma_\phi(p_0+q_0,|\vec p+ \vec q|)}{ (p^2 + 2 p \cdot q + q^2 - M^2)^2 + \Gamma_\phi^2(p_0+q_0,|\vec p+ \vec q|)} \nn \\
&& \times [\, \sign(p_0+q_0) - (\sign(q_0) + \sign(p_0))/2 ]^2.
\eea
Notice that if the two fermions are on-shell with $\sign(p_0)
\not=\sign(q_0)$, the momentum of the scalar $k^\mu=(\omega, \vec k)$ is
necessarily space-like, $\omega^2 - k^2 < 0$. The corresponding
Breit-Wigner width is close to equilibrium given by
\bea
\Gamma_\phi(\omega,k) &=& y^2 (\omega^2 - k^2)
\int_{(k-\omega)/2}^{(k+\omega)/2} \frac{dp}{8\pi k}
 n_\psi ( p) \nn \\
&=& \frac{y^2}{8\pi} \, \frac{ \omega^2 - k^2 }{\beta k} \,
\ln \left( \frac{1 + {\rm e}^{ \frac\beta2 (k-\omega)}}
{1 + {\rm e}^{ \frac\beta2 (k+\omega)}} \right) .
\eea
Integration over the energies yields
\bea
\label{Coff:Int}
{\cal C}^{\rm off}_\phi &=&  y^2 \int \frac{dp dq \sin \theta d\theta}{8\pi^4} p^2 q^2 
(1 - \cos \theta) \nn \\ 
&& \times \,  ( n_{\rm F}^{\rm eq}(-p) n_{\rm F}^{\rm eq}(q) n_{\rm B}^{\rm eq}(p-q) - n_{\rm F}^{\rm eq}(p) n_{\rm F}^{\rm eq}(-q) n_{\rm B}^{\rm eq}(q-p)   )\nn \\
&& \times \, \frac{\Gamma_\phi(|p-q|,\sqrt{p^2+q^2-2pq\cos\theta})}{ ( 2 pq (1 - \cos \theta) + M^2)^2 + \Gamma_\phi^2(|p-q|,\sqrt{p^2+q^2-2pq\cos\theta})}.
\eea
Here, we have used that $q_0$ and $p_0$ need to have opposite
signs. Notice that this does not contain the potentially harmful pole
in the bosonic distribution function, since the width $\Gamma$
vanishes in this case. A plot of ${\cal C}^{\rm off}_\phi$
is provided in Fig.~\ref{fig_Cs}.

Since the off-shell effects are proportional to the Breit-Wigner
width, they are proportional to $y^4$ and hence suppressed compared to
on-shell effects for small couplings. The resulting equilibration
rates given by $\gamma^Q$ as in eq.~(\ref{gamma_on}) and by the
eigenvalues of the system in eq.~(\ref{eq:off-shell}) for the choices
$y^2=1.0$ and $y^2=0.1$ are shown in Fig.~\ref{fig_gammas}.

It is again possible to make an estimate for the $M/T\gg1$ limit. The integrand
in eq.~(\ref{Coff:Int}) receives only relevant contributions
from $p,q\ll M$, such that we can replace the nominator by $M^4$.
Numerical evaluation then yields
\be
\label{Coff_high_M}
{\cal C}^{\rm off}_\phi\sim 0.126 \, y^4  \frac{T^8}{M^4}\,.
\ee
The suppression for large $M$ is not exponential, but given by the
Breit-Wigner width.  From Fig.~\ref{fig_gammas} it is however evident
that when $M/T$ is not too large and there is a sizable number of
scalar particles $S$ present in the heat bath, the on-shell
equilibration rate is dominating over the off-shell processes. Since
the on-shell rate is $\propto y^2$ and the off-shell rate $\propto
y^4$, this separation of scales becomes more pronounced for smaller
$y$.

\section{Discussion and Conclusions\label{sec_concl}}

We have calculated the various time-scales of chemical equilibration
in a theory with a Dirac fermion coupled to a scalar. For that
purpose, we have applied a gradient expansion and a linear response
ansatz to the Kadanoff-Baym equations. The different time-scales arise
hereby from processes that are driven either by on-shell or off-shell
effects. While on-shell equilibration is described in a quasi-particle
picture with $\delta$-functions accounting for the on-shell
conditions, we have shown that by including finite width effects, the
calculation can straightforwardly be generalized to determine the rate
of off-shell relaxation. We emphasize that already at leading loop
order in the 2PI effective action, a self-consistent solution to the
spectral and kinetic equations leads to off-shell chemical
equilibration. In addition, the gradient expansion is consistent with
off-shell chemical equilibration. Parametrically, we find for the time
scales of on- and off-shell chemical equilibration
\be
\gamma_{\rm on} \propto y^2, \quad
 \gamma_{\rm off} \propto y^4,
\ee
what explains the separation of time scales. In our model,
equilibration of charge results from on-shell processes, while there
exist two linear combinations of total particle numbers that only
relax through off-shell contributions.

For our discussion of the off-shell relaxation rate, we have focused
on the contributions to the collision integral where the scalar
particle is off-shell. Contributions of similar size (and same sign,
as can easily be checked) are expected from the domains of integration
where the fermions are off-shell. The numerical values presented
should hence not be taken at face value, but might be larger by a
factor $2-3$ in the full calculation, when taking the fermionic
off-shell effects into account. Since we find that the rate of
off-shell effects is of order $y^4$, three-loop contributions to the
effective action might also be equally important as the off-shell
effects arising from the two-loop effective action. However, the
analyses of Kadanoff-Baym equations in the literature we referred to
also take only the two-loop order of the effective action into
account. Hence, we neglected higher loop orders in the effective
action to facilitate a direct comparison. 

We would like to stress that our analysis does not only confirm the
common lore that off-shell effects can be important in certain models.
In the model at hand, full equilibration requires the violation of
eq.~(\ref{support}), which is a stronger condition than deviation from
the on-shell limit.  As a consequence of eq.~(\ref{support}) in
conjunction with the Breit-Wigner form of the spectral function, we
see from eq.~(\ref{Coff_high_M}) the decaying behavior of the
off-shell equilibration rate for large M. In contrast, if the spectral
function would be Gaussian with a width proportional to $y^2$,
off-shell equilibration would be exponentially suppressed.

The advantage of transport equations using the gradient expansion in
Wigner space is that they do not contain memory integrals and
therefore are much easier to solve numerically than the full
Kadanoff-Baym equations in coordinate space even if no further
approximations as e.g. the on-shell approximation or an ansatz in
terms of chemical potentials are used. Hence, it is important to
understand the range of applicability of transport equations. In
summary, our results indicate that the use of the gradient expansion
is justified to describe thermalization at rather late times (namely
after effects as prethermalization and damping are concluded) if the
off-shell effects are properly taken into account and coupling
constants are rather small. This is seen explicitly in the presented
analytical approach: In the on-shell approximation, the collision
terms vanish at all orders in gradients in some cases even though the
system is not in its true equilibrium state, while once including
off-shell effects, the system achieves full equilibration already at
first order in the gradient expansion. Hence, higher orders in the
gradient expansion are not decisive to describe full equilibration at
late times qualitatively in the present system.

\section*{Acknowledgments}
TK is supported by the EU FP6 Marie Curie Research \& Training Network
'UniverseNet' (MRTN-CT-2006-035863).  The work of BG is supported in
part by the U.S. Department of Energy contracts No. DE-FG02-08ER41531
and DE-FG02-95ER40896 and by the Wisconsin Alumni Research Foundation.
We thank the Institute for Nuclear Theory at the University of
Washington for its hospitality and the Department of Energy for
partial support during the completion of this work.

\end{document}